\documentclass[showkeys,eqsecnum,showpacs,prd,superscriptaddress]{revtex4}
\usepackage[dvips]{graphicx}
\usepackage{color}
\usepackage{amsmath}
\usepackage[dvips]{graphicx}
\begin{document}

\title{Partition Function of the Reissner-Nordstr\"om Black Hole}

\author{Jarmo M\"akel\"a} 

\email[Electronic address: ]{jarmo.makela@puv.fi}  
\affiliation{Vaasa University of Applied Sciences, Wolffintie 30, 65200 Vaasa, Finland}

\begin{abstract}

We consider a microscopic model of a stretched horizon of the Reissner-Nordstr\"om black hole. In our model the stretched horizon consists of discrete constituents. Using our model we obtain an explicit, analytic expression for the partition function of the hole. Our partition function implies, among other things, the Hawking effect, and provides it with a microscopic explanation as a phase transition taking place at the stretched horizon. The partition function also implies the Bekenstein-Hawking entropy law. The model and its consequences are similar to those obtained previously for the Schwarzschild black hole.
 
\end{abstract}


\maketitle

\section{Introduction}

 One of the key results of black hole physics is the {\it Bekenstein-Hawking entropy law}, which states that 
black hole has entropy, which is proportional to its event horizon area $A_H$: \cite{yy, kaa, koo}
\begin{equation}
S_{BH} = \frac{1}{4}\frac{k_Bc^3}{\hbar G}A_H.
\end{equation}
There have been numerous attempts to find a microscopc explanation to Eq. (1.1). In most cases those attempts 
have been based either on loop quantum gravity \cite{nee, vii} or string theory. \cite{kuu}

   In Ref. \cite{seite} Eq. (1.1) was obtained  for the Schwarzschild black hole by means of a microscopic model of a  specific stretched
horizon of the hole: The stretched horizon in question was chosen in such a way that the proper acceleration $a$ on 
the stretched horizon was always a constant, no matter what may happen to the Schwarzschild mass of the hole. The 
stretched horizon was assumed to consist of Planck-size constituents, each of them contributing to the stretched 
horizon an area, which is an integer times a constant. By means of that model an explicit, analytic expression was
obtained for the partition function of the Schwarzschild black hole from the point of view of an observer on its 
stretched horizon $a = constant$. The partition function implied, among other things, the Bekenstein-Hawking entropy
law of Eq. (1.1) in the special case, where the temperature of the hole agrees with its Hawking temperarure from the 
point of view of a distant observer at rest with respect to the hole. The model also implied the Hawking effect, and 
provided it with a microscopic explanation as a phase transition  taking place at the stretched horizon of the hole.

   In this paper we shall construct a similar microscopic model for the stretched horizon $a = constant$ of the 
Reissner-Nordstr\"om black hole, and obtain an explicit, analytic expression for the partition function of the hole. The results
obtainable from our partition function are analogous to those obtained in Ref. \cite{seite} for the Schwarzschild black hole. Unless
otherwise stated, we shall always use the natural units, where $\hbar = G = c = k_B = 4\pi\epsilon_0 = 1$.

\section{Stretched Horizon}

   The Reissner-Nordstr\"om black hole is an electrically charged, spherically symmetric black hole. The line element of spacetime
containing a Reissner-Nordstr\"om  black hole may be written in the spherical coordinates as:
\begin{equation}
ds^2 = -F(r,M,Q)\,dt^2 + \frac{dr^2}{F(r,M,Q)} + r^2\,d\theta^2 + r^2\sin^2\theta\,d\phi^2,
\end{equation}
where we have denoted:
\begin{equation}
F(r,M,Q) := 1 - \frac{2M}{r} + \frac{Q^2}{r^2}.
\end{equation}
In Eq. (2.2) $M$ is the mass of the hole and $Q$ is its electric charge. At the event horizon of the hole
\begin{equation}
r = r_+ := M + \sqrt{M^2 - Q^2},
\end{equation}
and one finds that the function $F = 0$ at the event horizon. For an extreme Reissner-Nordstr\"om black hole 
$M = \vert Q\vert$, and one finds that
\begin{equation}
F = (1 - \frac{M}{r})^2.
\end{equation}

   Consider now the proper acceleration of an observer outside of the event horizon of the Reissner-Nordstr\"om black hole.
The only non-vanishing component of the future pointing unit tangent vector field of the congruence of world lines of observers with constant cordinates $r$, $\theta$ and $\phi$ such that $r > r_+$ is
\begin{equation}
u^t = F^{-1/2},
\end{equation}
and hence it follows that the only non-vanishing component of the proper acceleration vector field
\begin{equation}
a^\mu := u^\alpha u^\mu_{;\alpha}
\end{equation}
of this congruence is
\begin{equation}
a^r = \frac{1}{2}\frac{\partial F}{\partial r}.
\end{equation}
The vector field $a^\mu$ is spacelike and it has a norm
\begin{equation}
a := \sqrt{a_\mu a^\mu} = \frac{1}{2}F^{-1/2}\frac{\partial F}{\partial r} 
= (1 - \frac{2M}{r} + \frac{Q^2}{r^2})^{-1/2}(\frac{M}{r^2} - \frac{Q¨2}{r^3}).
\end{equation}
This is the acceleration measured by an observer at rest with respect to the coordinates $r$, $\theta$ and $\phi$ for a particle in a free fall. As one may observe, far from the black hole the acceleration $a$ tends to its Newtonian value $\frac{M}{r^2}$.

   Our idea is to consider a spherical spacelike two-surface with constant $r$ and $t$ just outside of the event horizon as the stretched horizon of the Reissner-Nordstr\"om black hole. In the processes, where the mass $M$ and the electric charge $Q$  of the hole are changed, the radius $r$ of the stretched horizon is also changed, but in such a way that the proper acceleration $a$ given by Eq. (2.8) is kept as a constant. This means that if the mass $M$ and the electric charge $Q$ take on infinitesimal changes $dM$ and $dQ$, the radius $r$ of the stretched horizon will take on an infinitesimal change $dr$ such that
\begin{equation}
da = \frac{\partial a}{\partial r}\, dr + \frac{\partial a}{\partial M}\, dM + \frac{\partial a}{\partial Q}\, dQ = 0.
\end{equation}
Eq. (2.8) implies that this condition may be written as:
\begin{equation}
\left[\left(\frac{\partial F}{\partial r}\right)^2 - 2F\frac{\partial^2F}{\partial r^2}\right]\,dr
+ \left[\left(\frac{\partial F}{\partial r}\right)\left(\frac{\partial F}{\partial M}\right) 
- 2F\frac{\partial^2 F}{\partial r \partial M}\right]\,dM
+ \left[\left(\frac{\partial F}{\partial r}\right)\left(\frac{\partial F}{\partial Q}\right) 
-2F\frac{\partial^2 F}{\partial r\partial Q}\right]\,dQ = 0.
\end{equation}
When $r \longrightarrow r_+^+$, {\it i. e.} the stretched horizon tends to the event horizon of the hole, $F$ tends to zero.
Hence we see that very close to the event horizon we may write Eq. (2.10) which, in turn, follows from the condition (2.9), as:
\begin{equation}
\frac{\partial F}{\partial r}\,dr + \frac{\partial F}{\partial M}\,dM + \frac{\partial F}{\partial Q}\,dQ = 0.
\end{equation}
On the left hand side of this equation we have nothing else but the the total differential $dF$  of the function $F$. So we find that very close to the event horizon Eq. (2.9) implies:
\begin{equation}
dF = 0,
\end{equation}
which means that very close to the event horizon, where $F = 0$, constant $a$ implies constant $F$. Hence we may infer that a  stretched horizon originally close to the event horizon will stay close to the event horizon, when the mass $M$ and electric charge $Q$  of the hole are changed. In this sense our stretched horizon is well chosen.

   \section{The Model}

   In this paper we shall assume the the stretched horizon of the Reissner-Nordstr\"om black hole consists  of a finite number of discrete constituents, each of them contributing to the stretched horizon an area, which is  an integer times a constant. As a consequence, the stretched horizon area takes the form:
\begin{equation}
A = \gamma\ell_{Pl}^2(n_1 + n_2 +...+ n_N),
\end{equation}
where $n_1, n_2,..., n_N$ are non-negative integers, $\gamma$ is a pure number to be determined later, and
\begin{equation}
\ell_{Pl} := \sqrt{\frac{\hbar G}{c^3}} \approx 1.6\times 10^{-35} m
\end{equation}
is the Planck length. The integer $N$, which is assumed to be very large, is the number of the constituents of the stretched horizon. For a given black hole $N$ is assumed to be fixed. An immediate consequence of our model is that the area eigenvalues of the stretched horizon are of the form:
\begin{equation}
A_n = n\gamma\ell_{Pl}^2,
\end{equation}
where
\begin{equation}
n := n_1 + n_2 +...+ n_N.
\end{equation}
This means that the area spectrum of the stretched horizon of the Reissner-Nordstr\"om black hole has an equal spacing. Since the stretched horizon is assumed to lie very close to the event horizon of the hole, we may identify, for all practical purposes, the stretched horizon area of the Reissner-Nordstr\"om black hole with its event horizon area. The idea of an equally spaced event horizon area spectrum was raised by Bekenstein in 1974. \cite{kasi} Since then, equally spaced area spectra for the event horizons of black holes have been obtained by several authors on various grounds. \cite{ysi}

    Eq. (3.1) is assumed to hold, no matter, whether the Reissner-Nordstr\"om black hole is far from or close to extermality. The event horizon area of an extreme black hole with electric charge $Q$ is
\begin{equation}
A_{ext} = 4\pi Q^2,
\end{equation}
and therefore Eq. (3.1) implies that the eigenvalues $Q_n$ of the electric charge have a property:
\begin{equation}
Q_n^2 = \frac{\gamma}{4\pi}n,
\end{equation}
where $n$ is a non-negative integer. The charge spectrum of this form for the Reissner-Nordstr\"om black hole has been found, for instance, in Refs. \cite{kymmenen} and \cite{yytoo}. In Ref. \cite{yytoo} an interesting connection was pointed out between the charge spectrum given by Eq. (3.6) and Coleman's Big Fix mechanism, which suggests that the numerical values of the natural constants are fixed by an existence of baby universes and black holes. \cite{kaatoo}

   Since the Reissner-Nordstr\"om black hole has two independent, classical degrees of freedom, which are the mass $M$ and the electric charge $Q$ of the hole we associate with every constituent $j$ of the stretched horizon two non-negative
integers $s_j$ and $p_j$ such that
\begin{equation}
n_j = s_j + p_j.
\end{equation}
In our model the non-negative integers $s_j$ and $p_j$ are quantum numbers determining the quantum states of the constituent $j$ of the stretched horizon. Constituent $j$ is in vacuum, if $s_j = p_j = n_j= 0$; otherwise it is in an excited state. The electric chage $Q$ of the Reissner-Nordstr\"om black hole is assumed to be determined by the quantum numbers $p_j$ such that:
\begin{equation}
Q^2 = \frac{\gamma}{4\pi}(p_1 + p_2 +...+ p_N), 
\end{equation}
whereas for a black hole with fixed electric charge $Q$ the mass $M$ of the hole is determined by the quantum numbers $s_j$ such that
\begin{equation}
A - A_{ext} = \gamma\ell_{Pl}^2(s_1 + s_2 +...+ s_N).
\end{equation}
Since the event horizon area of the Reissner-Nordstr\"om black hole is $A = 4\pi r_+^2$, Eqs. (2.3) and (3.9) imply:
\begin{equation}
(M + \sqrt{M^2 - Q^2})^2 - Q^2 = \frac{\gamma}{4\pi}(s_1 + s_2 +...+ s_N).
\end{equation} 

\section{Energy}

   In the thermodymnamical considerations of all systems the concept of energy plays a central role. Our next aim is to obtain an expression for the energy of the Reissner-Nordstr\"om black hole from the point of view of an observer on its stretched horizon, where $a = constant$. If a particle with mass $dM$ and electric charge $dQ$ is carried through a two-sphere $r = constant$ $(r>r_+)$ in the Reissner-Nordstr\"om spacetime with a small relative speed, the energy carried through the two-sphere from the point of view of an observer at the spacelike infinity is
\begin{equation}
dE_\infty = dM - \frac{Q}{r}\,dQ.
\end{equation}
So we find that the energy carried through the two-sphere from the point of view of an observer on the two-sphere is:
\begin{equation}
dE = B(dM - \frac{Q}{r}\,dQ),
\end{equation}
where
\begin{equation}
B := F^{-1/2} = (1 - \frac{2M}{r} + \frac{Q^2}{r^2})^{-1/2}
\end{equation}
is the blue shift factor of the observer. Using Eq. (2.8) we may write Eq. (4.2) by means of the proper acceleration $a$ on the two-sphere as:
\begin{equation}
\left(\frac{\partial F}{\partial r}\right)\,dE = 2a(dM - \frac{Q}{r}\,dQ).
\end{equation}

   Eq. (4.4) is exact, and it always holds, no matter, whether our two-sphere is close to, or far from the event horizon of the hole. Close to the event horizon we may approximate the radius $r$ of the two-sphere by the radius $r_+$ of the event horizon of the Reissner-Nordstr\"om black hole. In this limit we find, by means of Eq. (2.3):
\begin{equation}
\frac{\partial F}{\partial r} = \frac{2M}{r_+^2} - \frac{2Q^2}{r_+^3} = \frac{2\sqrt{M^2 - Q^2}}{r_+^2}.
\end{equation}
We also find:
\begin{equation}
dr_+ =  \frac{\partial r_+}{\partial M}\,dM + \frac{\partial r_+}{\partial Q}\,dQ 
= \frac{r_+}{\sqrt{M^2 - Q^2}}\,dM  - \frac{Q}{\sqrt{M^2 - Q^2}}\,dQ,
\end{equation}
which means that close to the event horizon we have, in the leading approximation:
\begin{equation}
dM = \frac{\sqrt{M^2 - Q^2}}{r}\,dr + \frac{Q}{r}\,dQ,
\end{equation}
and Eq. (4.4) implies:
\begin{equation}
dE = ar\,dr.
\end{equation}
Because the area of the two-sphere is $A = 4\pi r^2$, we may write Eq. (4.8) in terms of the differential $dA = 8\pi r\,dr$ of the area as:
\begin{equation}
dE = \frac{a}{8\pi}\,dA.
\end{equation}
Hence it follows that the amount of energy carried through the stretched horizon, where $a = constant$ during the formation of the Reissner-Nordstr\"om black hole is
\begin{equation}
E = \frac{a}{8\pi}A.
\end{equation}
We shall use this expression as the energy of the Reissner-Nordstr\"om black hole from the point of view of an observer on its stretched horizon $a = constant$. In Ref. \cite{kootoo} a similar expression was obtained for the energy of the Kerr-Newman black hole. An advantage of our derivation of Eq. (4.10) is that it gives for the energy of the Reissner-Nordstr\"om black hole a natural interpretation as the energy carried through its stretched horizon during its formation.

 \section{Counting of States}

 The partition function of any system is, in general,
\begin{equation}
Z(\beta) := \sum_n g(E_n)e^{-\beta E_n},
\end{equation}
where we have summed over the energy eigenvalues $E_n$ of the system. $\beta$ is the inverse of the absolute temperature $T$ of the system, and $g(E_n)$ is the number of microscopic states associated with the same energy $E_n$ of the system. Eqs. (3.3), (3.4), (3.7) and (4.10) imply that the possible energies of the Reissner-Nordstr\"om black hole from the point of view of an observer on its stretched horizon, where the proper acceleration $a = constant$ are of the form:
\begin{equation}
E_n = n\gamma\frac{a}{8\pi},
\end{equation}
where
\begin{equation} 
n := s_1 + s_2 +...+ s_N + p_1 + p_2 +...+ p_N.
\end{equation}
We shall take the number $g(E_n)$ of the microscopic states associated with the same energy $E_n$ of the black hole to be the number of different combinations of non-vacuum area eigenstates of the constituents of the stretched horizon producing the same energy $E_n$. As a consequence, $g(E_n)$ is the number of possible ways of expressing the positive integer $n$ as a sum of at most $2N$ positive integers $s_j$ and $p_j$. More precisely, $g(E_n)$ is the number of ordered strings
\begin{equation}
{\cal S} := (s_1,s_2,...,s_{m_1};p_1,p_2,...,p_{m_2}),
\end{equation}
where $m_1, m_2 \le N$ such that at least one of the non-negative integers $m_1$ and $m_2$ is non-zero, and 
$s_1, s_2,..., s_{m_1}$ and $p_1, p_2,..., p_{m_2}$ are positive integers such that
\begin{equation}
s_1 + s_2 +...+s_{m_1} + p_1 + p_2 +...+ p_{m_2} = n.
\end{equation}
The number of such strings is
\begin{equation}
g(E_n) = \sum_{k=1}^{2N}\left(\begin{array}{cc}n-1\\k-1\end{array}\right),
\end{equation}
whenever $n \ge 2N$. In the special case, where $n = 2N$, we have:
\begin{equation}
g(E_n) = \sum_{k=1}^n\left(\begin{array}{cc}n-1\\k-1\end{array}\right) = 2^{n-1}.
\end{equation}
When $n < 2N$, $g(E_n)$ is simply the number of ways of writing a positive integer $n$ as a sum of positive integers, no m matter how many. That number is given by the right hand side of Eq. (5.7), and so we find that $g(E_n)$ is always given by Eq. (5.7), when $n \le 2N$.

\section{The Partition Function}

 Eqs. (5.1), (5.6) and (5.7) imply that the partition function of the Reissner-Nordstr\"om black hole from the point of view of an observer on the stretched horizon $a = constant$ is of the form:
\begin{equation}
Z(\beta) = Z_1(\beta) + Z_2(\beta),
\end{equation}
where
\begin{subequations}
\begin{eqnarray}
Z_1(\beta) &=& \frac{1}{2}\sum_{n=1}^{2N} 2^{(1-\beta T_C)n},\\
Z_2(\beta) &=& \sum_{n=2N+1}^\infty\left\lbrack\sum_{k=0}^{2N}
\left(\begin{array}{cc}n-1\\k\end{array}\right)2^{-n\beta T_C}\right\rbrack.
\end{eqnarray}
\end{subequations}
In these equations we have defined the temperature
\begin{equation}
T_C := \gamma\frac{a}{8\pi\ln 2}.
\end{equation}
The temperature $T_C$ plays an important role in the statistical and the thermodymical properties of the Reissner-Nordstr\"om black hole. Because of that we shall call $T_C$ as the {\it characteristic temperature} of the hole. 

  The expression in Eq. (6.1) for the partition function of the Reissner-Nordstr\"om black hole is almost identical to the partition function obtained in Ref. \cite{seite} for the Schwarzschild black hole; the only difference is that for the Schwarzschild black hole we replace $2N$ in Eqs. (6.2a) and (6.2b) by $N$. Replacing $N$ by $2N$ in Eqs. (155) and (156) in Ref. \cite{seite} we find that the partition function of the Reissner-Nordstr\"om black hole may be calculated explicitly, and it takes the form:
\begin{equation}
Z(\beta) = \frac{1}{2^{\beta T_C} - 2}\left[1 - \left(\frac{1}{2^{\beta T_C} - 1}\right)^{2N+1}\right],
\end{equation}
when $\beta T_C \ne 1$ and 
\begin{equation}
Z(\beta) = 2N + 1,
\end{equation}
when $\beta T_C = 1$.

\section{Energy, Entropy and the Hawking Effect}

      Since the partition function of the Reissner-Nordstr\"om black hole in Eq. (6.4) is almost identical to the partition function obtained for the Schwarzschild black hole in Ref. \cite{seite}, the properties of the average energy
\begin{equation}
 E(\beta) = -\frac{\partial}{\partial \beta}\ln Z(\beta)
\end{equation}
and the entropy
\begin{equation}
S(\beta) = \beta E(\beta) + \ln Z(\beta)
\end{equation}
of the Reissner-Nordstr\"om black hole are very similar to those of the Schwarzschild black hole. One finds that the average energy per a constituent
\begin{equation}
{\bar E}(\beta) := \frac{E(\beta)}{N}
\end{equation}
takes in the large $N$ limit in the leading approximation the form:
\begin{equation}
{\bar E}(\beta) = {\bar E}_1(\beta) + {\bar E}_2(\beta),
\end{equation}
where:
\begin{subequations}
\begin{eqnarray}
{\bar E}_1(\beta) :&=& \frac{1}{N}\frac{2^{\beta T_C}}{2^{\beta T_C} - 2}T_C\ln 2,\\
{\bar E}_2(\beta) :&=&-\frac{2^{\beta T_C}}{(2^{\beta T_C} - 1)^{2N+2} - 2^{\beta T_C} + 1}2T_C\ln 2.
\end{eqnarray}
\end{subequations}
Eqs. (7.3) and (7.4) imply that when $\beta T_C > 1$, which means that $T < T_C$, we have:
\begin{equation}
\lim_{N\rightarrow\infty}{\bar E}(\beta) = 0,
\end{equation}
which states that the constituents of the stretched horizon are effectively in vacuum, when $T < T_C$, whereas if 
$T > T_C$, we have:
\begin{equation}
\lim_{T\rightarrow T_C^+}{\bar E}(\beta) = 4T_C\ln 2.
\end{equation}
in the large $N$ limit. Hence we observe that at the characteristic temperature $T_C$ the Reissner-Nordstr\"om black hole performs a {\it phase transition}, and the latent heat per a constituent associated with this phase transition is:
\begin{equation}
{\bar L} = 4T_C\ln 2.
\end{equation}
Using Eqs. (3.1), (4.10) and (7.8) one concludes that after the phase transition has been completed, we have:
\begin{equation}
{\bar n} = 4,
\end{equation}
where
\begin{equation}
{\bar n} := \frac{n_1 + n_2 +... + n_N}{N}
\end{equation}
is the average of the quantum numbers $n_1,..., n_N$ defined in Eq. (3.1). This means that during the phase transition the constituents of the stretched horizon jump, in average, from the vacuum to the fourth excited states.

    Since the constituents are effectively in vacuum, when $T < T_C$, the characteristic temperature $T_C$ is the lowest possible temperature the Reissner-Nordstr\"om black hole may have from the point of view of an observer on its stretched horizon: When $T < T_C$ and the constituents of the stretched horizon are in vacuum, and there is no black hole. Using Eqs. (2.8) and (6.3) we find that in the leading approximation close to the event horizon we have:
\begin{equation}
T_C = \gamma B\frac{\kappa}{8\pi\ln 2},
\end{equation}
where $B$ is the blue shift factor of the observer, defined in Eq. (4.3),  and
\begin{equation}
\kappa := \frac{\sqrt{M^2 - Q^2}}{r_+^2}
\end{equation}
is the srface gravity of the Reissner-Nordstr\"om hole at its event horizon. The Tolman relation \cite{neetoo} implies that from the point of view of a distant observer the temperature of the hole is
\begin{equation}
T_\infty = \gamma\frac{\kappa}{8\pi\ln 2},
\end{equation}
which agrees with the {\it Hawking temperature}  \cite{nee}
\begin{equation}
T_H := \frac{\kappa}{2\pi},
\end{equation}
if we put
\begin{equation}
\gamma = 4\ln 2.
\end{equation}
Hence we have obtained the {\it Hawking effect} from our model: The Reissner-Nordstr\"om black hole has a certain minimum temperature, which agrees with its Hawking temperature from the point of view of a distant observer. When the temperature of the environment of the hole is dropped below its Hawking temperature, the hole begins to radiate. When the hole radiates, it performs a phase transition, where the constituents of its stretched horizon descend, in average, from the fourth excited states to the vacuum, and radiation is emitted. Finally, all constituents (except possibly one) have reached the vacuum, and the hole has evaporated away.

  Calculations similar to those performed in Ref. \cite{seite} for the Schwarzschild black hole indicate that the entropy of the Reissner-Nordstr\"om black hole may be written in terms of its stretched horizon area $A$ as:
\begin{equation}
S(A)  = \frac{1}{4\ln 2}\frac{k_B c^3}{\hbar G}A\ln\left(\frac{2A}{2A - A_{crit}}\right) 
+ 2N_B\ln\left(\frac{2A - A_{crit}}{A_{crit}}\right),
\end{equation}
whenever $A > A_{crit}$, where the critical area
\begin{equation}
A_{crit} := 16N\ell_{Pl}^2\ln 2
\end{equation}
is the stretched horizon area, when $T\longrightarrow T_C^+$. Putting $A =  A_{crit}$ in Eq. (7.16) one finds:
\begin{equation}
S(A) = \frac{1}{4}\frac{k_B c^3}{\hbar G}A,
\end{equation}
which is exactly the Bekenstein-Hawking entropy law of Eq. (1.1). One may show that Eq. (7.18) holds, whenever 
$A \le A_{crit}$. Hence we have obtained the Bekenstein-Hawking etropy law from our model in the special case, where the temperature of the hole from the point of vew of a distant observer agrees with its Hawing temperature. However, there is nothing to prevent the temperature of the hole from being higher than its Hawking temperature. When the temperature of the hole exceeds its Hawking temperature, the entropy of the hole differs from its Bekenstein-Hawking entropy.

\section{Discussion}

 In this paper we have considered the thermodynamics and the statistical physics of the Reissner-Nordstr\"om black hole from the point of view of an observer on a stretched horizon, where the proper acceleration of the observer is a constant, no matter what may happen to the mass $M$ and the electric charge $Q$ of the hole. We assumed that the stretched horizon consists of discrete constituents, each of them contributing to the stretched horizon an area, which is an integer times a constant. In this sense our model was similar to the one previously constructed for the Schwarzschild black hole. \cite{seite} The quantum states of the constituents $j$ of the stretched horizon were determined by two quantum numbers $p_j$ and $s_j$, of which the sum of the quantum numbers $p_j$ determined the electric charge of the hole, and the sum of the quantum numbers $s_j$ determined the mass of the hole under presence of a given electric charge. Using our model we managed to obtain for the partition function of the Reissner Nordstr\"om  black hole an explicit, analytic expression, which is almost identical to the partition function previously obtained for the Schwarzschild black hole. \cite{seite} Our partition function implied, among other things, that the Reissner-Nordstr\"om black hole has a certain minimum temperature which, from the point of view of a distant observer, agrees with its Hawking temperature. When the temperature of the hole from the point of view of a distant observer agrees with its Hawking temperature, the hole performs a phase transition, where the constituents of its stretched horizon descend, in average, from the fourth excited states to the vacuum, and radiation is emitted. In this sense our model implies the Hawking effect for the Reissner-Nordstr\"om black hole. Our partition function also implied the Bekenstein-Hawking entropy law for the Reissner-Nordstr\"om black hole in the special case, where the temperarture of the hole agrees with its Hawking temperature. However, if the temperature of the hole is higher than its Hawking temperature, the entropy of the hole will differ from its Bekenstein-Hawking entropy. 

     Even though our model meets with some success in the sense that it reproduces for the Reissner-Nordstr\"om black hole the results previously obtained for the Schwarzschild black hole and provides a microscopic explanation to the Hawking effect and the Bekenstein-Hawking entropy law, the model involves some problems as well. One of them is the curious result implied by Eq. (3.8) that the square $Q^2$ of the electric charge of the Reissner-Nordstr\"om black hole, instead of the electric charge $Q$ itself, has an equally spaced spectrum. This is not necessarily a problem, because there is not any compelling reason why the charge spectrum of a black hole should be similar to that of ordinary matter. Anyway, if one does not like the charge spectrum implied by Eq. (3.8), one may attempt to consider the electric charge of the Reissner-Nordstr\"om black hole as a mere external parameter of the model. In that case the quantum numbers $p_1, p_2,..., p_N$ in Eq. (3.8) are all constants, and the square of the electric charge is fixed to be:
\begin{equation}
Q^2 = \frac{\gamma}{4 \pi}p,
\end{equation}
where $p := p_1 + p_2 +...+ p_N$. The quantum states of the hole are solely determined by the quantum numbers $s_1, s_2,..., s_N$ such that
\begin{equation}
s_1 + s_2 +...+ s_N + p = n.
\end{equation}
As a consequence, the number of microscopic states associated with the energy $E_n$ is
\begin{equation}
g(E_n) = \sum_{k=1}^{N}\left(\begin{array}{cc}n-p-1\\k-1\end{array}\right),
\end{equation}
when $n - p > N$, and
\begin{equation}
g(E_n) = 2^{n-p-1},
\end{equation}
when $n - p \le N$. The partition function turns out to be
\begin{equation}
Z(\beta) = \frac{2^{-p\beta T_C}}{2^{\beta T_C} - 2}\left[1 - \left(\frac{1}{2^{\beta T_C} - 1}\right)^{N+1}\right],
\end{equation}
when $\beta T_C \ne 1$. When $\beta T_C = 1$, one finds that $Z(\beta) = (N+1)2^{-p}$. So we observe that our partition function depends on the electric charge of the hole through Eq. (8.1). Using the new partition function one may again obtain the Hawking effect, but the expression for the entropy of the hole takes the form:
\begin{equation}
S(A) = \frac{1}{4\ln 2}\frac{k_Bc^3}{\hbar G}\Delta A\ln\left(\frac{2\Delta A}{2\Delta A - A_{crit}}\right) 
+ Nk_B\ln\left(\frac{2\Delta A - A_{crit}}{A_{crit}}\right),
\end{equation}
where 
\begin{equation}
\Delta A := A - A_{ext},
\end{equation}
and 
\begin{equation}
A_{crit} : = 8N\ell_{Pl}^2\ln 2.
\end{equation}
In the limit, where $\Delta A\longrightarrow A_{crit}^+$, which means that $T \longrightarrow T_C^+$, the entropy takes the form:
\begin{equation}
S(A) = \frac{1}{4}\frac{k_b c^3}{\hbar G}(A - A_{ext}).
\end{equation}
Hence we find that the enropy of an extreme black hole vanishes, when the electric charge is treated as a mere external parameter. However, if the electric charge is quantized as in Eq. (3.8), the entropy of the hole is given by Eq. (1.1), no matter, whether the hole is extreme or not.

\end{document}